\ifx\LaTeXe\undefined
\documentstyle[12pt,epsfig,latexsym]{article}

\else  
\documentclass[11pt]{article}
\usepackage{a4}
\usepackage[dvips]{graphicx}
\usepackage{latexsym}
\usepackage{euscript}
\usepackage{amssymb}
\usepackage{epsfig}

\fi
\newcommand{\Tr}{\mbox{Tr}\;}

\newcommand{\simge}{\ \lower-
1.2pt\vbox{\hbox{\rlap{$>$}\lower5pt
\vbox{\hbox{$\sim$}}}}\ }

\newcommand{\bc}{\begin{center}}
\newcommand{\ec}{\end{center}}
\newcommand{\be}{\begin{equation}}
\newcommand{\ee}{\end{equation}}
\newcommand{\ba}{\begin{eqnarray}}
\newcommand{\ea}{\end{eqnarray}}
\newcommand{\brr}{\begin{array}}
\newcommand{\err}{\end{array}}
\newcommand{\Nf}{N_{\rm f}}
\def\lsi{\raise0.3ex\hbox{$<$\kern-0.75em\raise-1.1ex\hbox{$\sim$}}}
\def\gsi{\raise0.3ex\hbox{$>$\kern-0.75em\raise-1.1ex\hbox{$\sim$}}}

\newcommand{\gsim}{\mathop{\gsi}}

\begin{document}
\pagestyle{empty}
\vspace{-0.6in}
\begin{flushright}
\vskip 0.2cm
BI-TP 2004/03\\
CPT-2003/P.4622\\
DESY 04-009\\
FTUV-04-0203\\
IFIC/04-04\\
MPP-2004-6\\
\end{flushright}
\vskip 2.0cm
\centerline{\Large {\bf{Low-energy couplings of QCD from current}}}
\vskip 0.25cm
\centerline{\Large {\bf{correlators near the chiral limit}}}
\vskip 0.75cm
\centerline{L. Giusti$^{a}$, P. Hern\'andez$^{b}$, M. Laine$^{c}$, 
            P. Weisz$^{d}$, H. Wittig$^{e}$}
\vskip 0.5cm
\centerline{$^a$ Centre de Physique Th\'eorique, CNRS Luminy, Case 907, 
            F-13288 Marseille, France}
\vskip 0.1cm
\centerline{$^b$ Dep. de F\'isica Te\`orica, Univ. de Val\`encia, 
            E-46100 Burjassot, Spain}
\vskip 0.1cm
\centerline{$^c$ Faculty of Physics, University of Bielefeld, D-33501 Bielefeld, Germany}
\vskip 0.1cm
\centerline {$^d$ Max-Planck-Institut f\"ur Physik, F\"ohringer Ring 6, D-80805 Munich, Germany}
\vskip 0.1cm
\centerline{$^e$ DESY, Theory Group, Notkestrasse 85, D-22603 Hamburg, Germany}
\vskip 1.5cm
\begin{abstract}
\noindent We investigate a new numerical procedure to compute fermionic 
correlation functions at very small quark masses. 
Large statistical fluctuations, due to the presence of 
local ``bumps'' in the wave functions associated with the low-lying eigenmodes 
of the Dirac operator, are reduced by an exact low-mode averaging.
To demonstrate the feasibility of the technique, we compute the 
two-point correlator of the left-handed vector current with 
Neuberger fermions in the quenched approximation, for lattices 
with a linear extent of $L\approx 1.5$~fm, a lattice 
spacing $a\approx 0.09$~fm, and quark masses down to the 
$\epsilon$-regime. By matching the results with the corresponding 
(quenched) chiral perturbation theory expressions, an estimate of 
(quenched) low-energy constants can be obtained. We find agreement 
between the quenched values of $F$ extrapolated from the 
$p$-regime and extracted in the $\epsilon$-regime. 
\end{abstract}
\vskip 1.5cm
\noindent March 2004
\vfill
\pagestyle{empty}\clearpage
\setcounter{page}{1}
\pagestyle{plain}
\newpage
\pagestyle{plain} \setcounter{page}{1}

\newpage

\section{Introduction}
One of the most important challenges for lattice QCD is the simulation of
quarks with masses light enough to reach kinematical regions where chiral 
perturbation theory (ChPT) \cite{Weinberg:1978kz,Gasser:1983yg}
can be verified and safely applied. Despite valuable efforts 
in the infinite-volume limit \cite{Butler:em}-\cite{Gattringer:2003qx}, 
it is still unclear how small the quark masses need to be in order to reach  
the chiral region (for recent developments, see \cite{Bernard:2002yk,Dong:2003im}). 
More recently the $\epsilon$-regime of QCD \cite{Gasser:1987ah,N}, in 
which $m\rightarrow 0$ at finite volume $V$, is being attacked 
on the 
lattice~\cite{Damgaard:1999tk,condensate,epsilon,Hasenfratz:2002rp,Bietenholz:2003mi,Giusti:2003gf,Giusti:2003iq}.
Difficulties to reach these kinematical corners arise because most of 
the numerical techniques currently used on the lattice become 
inefficient when the quark masses approach the chiral limit.\\ 
\indent In the presence of spontaneous chiral symmetry breaking, chiral 
perturbation theory suggests that the low-lying eigenvalues of the 
massless QCD Dirac operator scale proportionally to $(\Sigma V)^{-1}$, where 
$\Sigma$ is the bare quark condensate and $V$ is the lattice volume 
\cite{Gasser:1987ah,Leutwyler:1992yt}. In addition random matrix 
theory \cite{Shuryak:1992pi,Nishigaki:1998is} 
(and to some extent quenched ChPT \cite{Akemann:2003tv}) predicts the 
distribution of each individual eigenvalue as a function of $\Sigma V$.
Comparisons of 
these predictions with numerical data obtained on the lattice in quenched 
QCD have been reported by several collaborations
\cite{Edwards:1999ra,Damgaard:1999tk,Bietenholz:2003mi,Giusti:2003gf}.
Only recently has an extensive study of this issue, performed at 
several volumes and lattice spacings, shown a detailed 
agreement between some of these predictions and quenched QCD 
for volumes larger than about $5$~fm$^4$ \cite{Giusti:2003gf}.
No predictions are available for the properties of the  
corresponding wave functions
(for a recent compilation of numerical results, see \cite{Edwards:2001ei}).
It is an empirical observation, though, that they can develop local 
``bumps'' with a non-negligible probability \cite{Giusti:2003gf}.\\ 
\indent It is then conceivable to exploit such basic properties of QCD as suggested by 
ChPT to develop exact and generic numerical algorithms which remain efficient 
when the quark masses approach the chiral regime. For instance, due to the 
scaling of the low-lying eigenvalues with $V$, the computation  
of the fermion propagator with standard techniques becomes very demanding 
when the quark masses approach 
the chiral limit, since the Dirac operator tends to be strongly ill-conditioned. 
The slow-down can be dramatically reduced by subtracting a few of the lowest 
modes and treating them separately~\cite{Giusti:2002sm}.\\ 
\indent In this paper we propose a 
new technique\footnote{During the writing of this paper, T. DeGrand and 
S. Schaefer applied a technique similar to the one presented in this 
work to reduce statistical noise in the computation of 
two-point correlation functions of bilinear operators \cite{DeGrand:2004qw}. 
A similar idea has also been sketched by R.G. Edwards to reduce noise 
in the computation of singlet correlation functions 
\cite{Edwards:2001ei}.} to compute QCD 
fermionic correlation functions at very small quark masses on the lattice. 
An exact low-mode averaging procedure is introduced to reduce large statistical 
fluctuations induced by the presence of ``bumps'' in the wave functions 
associated with the low-lying eigenvalues of the Dirac operator
which happen to occur where the fermionic operators are localized.
The feasibility of the technique is proven by computing the 
correlator of two left-handed vector currents in quenched 
QCD with Neuberger fermions \cite{Ginsparg:1982bj}-\cite{Hernandez:1998et}.
We find that the variance of the estimate is 
markedly reduced with respect to the one with standard techniques 
when the quark mass is 
decreased, and the $\epsilon$-regime can be safely 
reached in all topological sectors.\\
\indent The use of the quenched approximation mainly serves to test our
method, which we expect will be effective for simulations of full QCD
as well. It is known that the quenched theory is afflicted with
several problems. In particular, the removal of the fermion
determinant renders the theory non-unitary. An effective low-energy
description of the quenched theory is formally obtained if an
additional expansion in $1/N_c$, where $N_c$ is the number of colors,
is carried out together with the usual one in quark masses and
momenta. The resulting so-called quenched ChPT 
\cite{Bernard:1992mk,Sharpe:1992ft} leads to
infrared divergences in certain correlation functions. These
divergences reflect, at least partially, the sickness of quenched
QCD. Here we adopt the pragmatic assumption that -- despite the fact that
it is not an asymptotic expansion of quenched QCD (at fixed $N_c$) --
quenched ChPT describes the low-energy regime of quenched QCD in
certain ranges of kinematical scales, where correlation functions can
be parameterized in terms of effective coupling constants,            
the latter being defined as the couplings which appear in the         
Lagrangian of the effective theory. With this
assumption in mind we compare the predictions of quenched ChPT with
the numerical results obtained in the kinematically accessible ranges
in the $p$- and $\epsilon$-regimes for the correlator of two
left-handed currents. Thereby we estimate values of $F$ and $\alpha_5$
(to be defined below), which, according to our working assumption, are
then identified with quenched versions of the physical LECs.\\
\indent It is interesting to note that the correlator 
of two left-handed vector currents is free from zero-mode 
contributions and, at fixed volume, remains 
finite when the quark mass $m\rightarrow 0$. This is in contrast to the 
correlators of two scalar or pseudoscalar densities in sectors of 
non-zero topological charge at finite volume. They 
develop $1/m^2$ divergences with residues given by correlation 
functions of zero-mode wave functions \cite{Giusti:2003iq}. 
Since in both cases the leading non-trivial behaviour is governed by 
$F$, these two different types of correlators offer independent determinations of 
this coupling.\\
\indent The paper is organized as follows: in Sect.~\ref{sec:ChPT} 
we collect the formul\ae~for the 
two-point correlation function of the left-handed vector current in chiral perturbation 
theory, in Sect.~\ref{sec:lowm} we describe the low-mode 
averaging procedure for fermionic correlation
functions, and in Sect.~\ref{sec:numeric} we report the details of the simulations 
we have performed in quenched QCD, as well as the results obtained. We conclude in 
Sect.~\ref{sec:concl}. In the first Appendix we provide more details of our notations,
and in the second we collect further useful formul\ae~obtained in ChPT at finite
volume.  

\section{Left-handed current correlator in ChPT}\label{sec:ChPT}
We start by considering the physical, unquenched QCD.
The Euclidean Lagrangian of the chiral effective theory for this case reads, 
at leading order in the momentum expansion, 
\be
  {\cal L}=\frac{F^2}{4}\,
  \Tr\left\{\partial_{\mu}U^{\dagger}\partial_{\mu}U\right\}
  -\frac{\Sigma}{2} \,\Tr\left\{e^{i\theta/\Nf} UM+M^{\dagger}U^{\dagger} 
e^{-i\theta/\Nf}\right\}\, ,
\ee 
where $U \in $ SU($\Nf$) is the meson field, $\Nf=3$,
$M = \mathop{\mbox{diag}}(m_u,m_d,m_s)$ the quark mass matrix, 
$\theta$ the vacuum angle, and, in the chiral limit, 
$F$ equals the pseudoscalar decay constant 
and $\Sigma$ the chiral condensate \footnote{%
   In this section we use continuum notation.}.
At the next-to-leading order (NLO)
in the momentum expansion, additional operators
appear in the effective Lagrangian, 
with the associated LECs $\alpha_1,\alpha_2,...$~\cite{Gasser:1983yg}~\footnote{%
  We adopt the convention of \cite{Heitger:2000ay} where 
  $\alpha_i = 8 (4\pi)^2 L_i$, with $L_i$ as defined in~\cite{Gasser:1983yg}. 
  }.
In the following we consider only the case of a degenerate mass matrix, 
i.e.\ $m_u = m_d = m_s \equiv m$.\\
\indent The LECs can be determined by computing suitable QCD correlation functions
on the lattice at small masses and momenta, and by comparing the results 
with the parameterization given by ChPT. In this paper we are interested
in the two-point correlation function~\cite{Giusti:2002sm,Hernandez:2002ds}
\be
{\cal C}^{ab}(t) = \int {\rm d}^3 x \, \langle {\cal J}^a_0(x) {\cal J}^b_0(0) \rangle 
\ee
of the left-handed charge density ${\cal J}^a_0(x)$, where $t=x_0$. 
In the effective theory, at the leading order in momentum 
expansion, it reads 
\be
  {\cal J}^a_{0}=\frac{F^2}{2}\,
  \Tr\left\{\mbox{T}^a U\partial_{0}U^{\dagger}\right\}
\ee 
where $\mbox{T}^a$ is a traceless generator of SU($\Nf$), 
acting on flavor indices, and on the side of QCD we assumed the
normalization of Eq.~(\ref{eq:j0qcd}) below.\\
\indent A kinematical range of scales which is suitable for extracting the LECs
in a finite box of volume $V = T L^3$, is the so-called $p$-regime. 
It is defined by the constraints $M_P L \gsim 1$ and $M_P \ll 4 \pi F$, where $M_P$ is 
the pseudoscalar meson mass. Writing 
$\mathcal{C}^{ab}(t) \equiv \Tr [\mbox{T}^a \mbox{T}^b]\, \mathcal{C}(t)$, the 
next-to-leading order (NLO) finite-volume prediction can 
be expressed as~\cite{Hansen:1990un}
\be
 \mathcal{C}(t) 
 = \frac{1}{2} M_P^V (F_P^V)^2 \frac{\cosh\Bigl[ (T/2 - |t| )M_P^V\Bigr]}
   {2 \sinh\Bigl[ T M_P^V/2\Bigr]} 
 - \frac{\Nf}{2} \frac{{\rm d} g_1}{{\rm d}T}\, , \label{eq:Ct_p}
\ee
where 
\ba
 F_P^V & \equiv & F_P \biggl( 
 1 - \frac{\Nf g_1}{2 F_P^2} \biggr)  \; ,\label{eq:FPV} \\
F_P & =	& F 
 \biggl[ 
 1 - \frac{\Nf G(M^2)}{2 F^2} + \frac{M^2}{2 (4\pi F)^2}
 \Bigl( \Nf \alpha_4 + \alpha_5 \Bigr)
 \biggr] \;,
\ea
and $M^2=2m\Sigma/F^2$. The effective finite-volume meson mass $M_P^V$ and the 
functions $g_1$  and $G(M^2)$ 
are reported in Appendix B. Finite-volume effects
are exponentially small if $M_P L \gg 1$, and we can set $g_1 = 0$.\\
\indent A less explored kinematical region of QCD, where the LECs can be extracted, 
is the so-called  $\epsilon$-regime, where $M_P L \ll 1$ and the linear extent 
of the box is such that $ 4 \pi FL\gg 1$ \cite{Gasser:1987ah,N}. 
In this regime topology plays an important r\^ole \cite{Leutwyler:1992yt}, and 
for fixed topological charge $\nu$ \cite{Hansen:1990un,Damgaard:2002qe,Hernandez:2002ds}
\be\label{eq:jljl-e}
 \mathcal{C}(t) 
 = \frac{F^2}{2 T}
 \biggl[
 1 + \frac{\Nf}{F^2}\biggl(
 \frac{\beta_1}{\sqrt{V}} - \frac{T^2 k_{00}}{V} \biggr)
 + \frac{2 \mu T^2}{F^2 V} \sigma_\nu(\mu) h_1\Bigl(\frac{t}{T}\Bigr) 
 \biggr] 
 \;, \label{eq:Ct_eps}
\ee
where $\mu\equiv m\Sigma V$ and 
\be
h_1\left(\frac{t}{T}\right) = 
\frac{1}{2}\left[\Big(\Big|\frac{t}{T}\Big|-\frac{1}{2}\Big)^2 
-\frac{1}{12} \right]\, .
\ee
The constants $\beta_1$ and $k_{00}$ are related to 
the (dimensionally regularized) value of 
\be
 \bar G(0) = \frac{1}{V} 
 \sum_{n \in \mbox{\raisebox{-0.2ex}
 {$\scriptstyle\mathsf{Z\hspace*{-1.3mm}Z}^4$}}} 
 \Bigl(1 - \delta^{(4)}_{n,0} \Bigr) \frac{1}{p^2} 
 \;, \quad
 p = 2\pi\Bigl( \frac{n_0}{T}, \frac{\vec{n}}{L} \Bigr)
 \;,
 \label{eq:Gx}
\ee
by
\be
 \bar G(0) \equiv -\frac{\beta_1}{\sqrt{V}} \;, \quad
 T \frac{{\rm d}}{{\rm d} T} \bar G(0) \equiv \frac{T^2 k_{00}}{V} 
 \;.
\ee
Furthermore $\sigma_\nu(\mu) \equiv {\Nf}^{-1}  {\rm d}\{ \ln \det[I_{\nu+j-i}(\mu)] \} / 
{\rm d} \mu$, 
where the determinant is taken over an $\Nf \times \Nf$ matrix, whose 
matrix element $(i,j)$ is the modified Bessel function $I_{\nu+j-i}$ 
\cite{Brower:vt,Leutwyler:1992yt}.\\
\indent In ChPT there is a well-defined prescription to compute correlation 
functions at fixed topology, and here we assume that also in QCD they 
have a well-defined meaning in the continuum limit at non-zero physical distances. 
Although plausible, this is a non-trivial 
dynamical issue and to pose precise questions we must 
introduce an ultraviolet regularization. We here adopt a 
lattice regularization with fermions discretized {\it \`a la} Neuberger \cite{Neuberger}.
The massless Dirac operator obeys the Ginsparg-Wilson (GW) relation
and therefore preserves an exact chiral symmetry. The 
topological index assigned to a configuration is 
$\nu=n_+-n_-$, where $n_+$ ($n_-$) are the numbers of zero-modes 
of $D$ with positive (negative) chirality. Our working hypothesis is 
that correlators of composite operators at non-zero physical distance
 have a continuum limit in any given sector of index $\nu$ 
independent of the particular choice of $D$  
\footnote{Since the space of lattice gauge fields is connected, 
  different choices of $D$ possibly lead to different assignments
  of index for a given configuration.}.
Some recent numerical evidence (in the quenched approximation) 
consistent with this scenario can be found, e.g. in 
Refs.~\cite{Giusti:2003gf,DelDebbio:2003rn}.\\
\indent In quenched ChPT the flavor singlet field cannot be integrated out
and therefore additional coupling constants appear in the chiral Lagrangian
\cite{Bernard:1992mk,Sharpe:1992ft}. 
In particular the singlet mass parameter $m^2_0/2N_c$ is dimensionful and, even
if suppressed by large-$N_c$ counting, cannot be tuned by changing the 
kinematical conditions. Consequently, the standard chiral expansion is expected to 
be useful only in a window of Euclidean momenta $q^2$, 
where $m_0^2 / 2 N_c \ll q^2 \ll (4\pi F)^{2}$.\\
\indent In the quenched theory, Eq.~(\ref{eq:Ct_p}) remains the same,
apart from the omission of the last constant term $\propto\Nf$, 
but at NLO the interpretation of the parameters in terms of those 
of quenched ChPT changes \cite{Colangelo:1997ch} \footnote{%
Analogous LECs in ChPT and quenched ChPT are indicated with the same
symbols, since they can be clearly distinguished from the context.}. In particular,
the parameter $F_P^V$ is volume-independent,
\be
 F_P^V = F_P =  F \biggl[ 
 1 + \frac{M^2}{2(4\pi F)^2} \alpha_5 \biggr]\, ,  \label{eq:FPq} \\
\ee
where the LEC $\alpha_5$ is finite at this order. 
In the $\epsilon$-regime, on the other hand, the correlator of two left-handed vector charges 
is modified to be \cite{Damgaard:2002qe,Hernandez:2002ds}
\be\label{eq:eps_quenched}
{\cal C}(t) =  \frac{F^{2} }{2 T} \left\{ 1 + 
\frac{2\mu T^2}{F^{2} V} \sigma_\nu(\mu) h_1\Big(\frac{t}{T}\Big)\right\}\, ,
\ee
where in this case \cite{dotv}
\be
\sigma_\nu(\mu) = \mu \Big\{I_\nu(\mu) K_\nu(\mu) + I_{\nu+1}(\mu) 
K_{\nu-1}(\mu) \Big\} + \frac{\nu}{\mu}\, ,
\ee
and $I_\nu$ and $K_\nu$ are modified Bessel functions. 

\section{Low-mode averaging}\label{sec:lowm}
\begin{figure}[thb]
\begin{center}
\begin{tabular}{cc}
\mbox{\epsfig{file=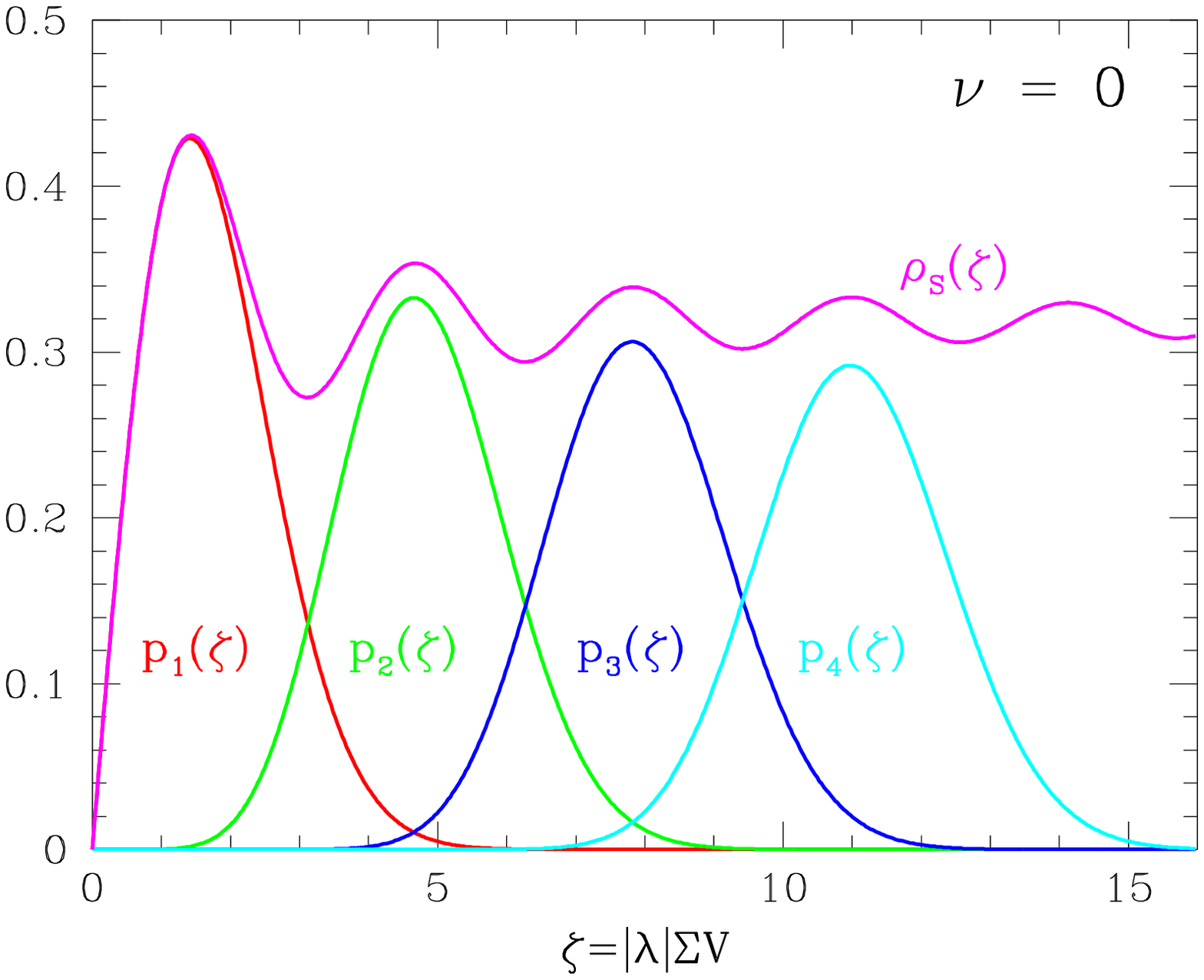,height=6.0cm,width=6.5cm,angle=0}} &
\mbox{\epsfig{file=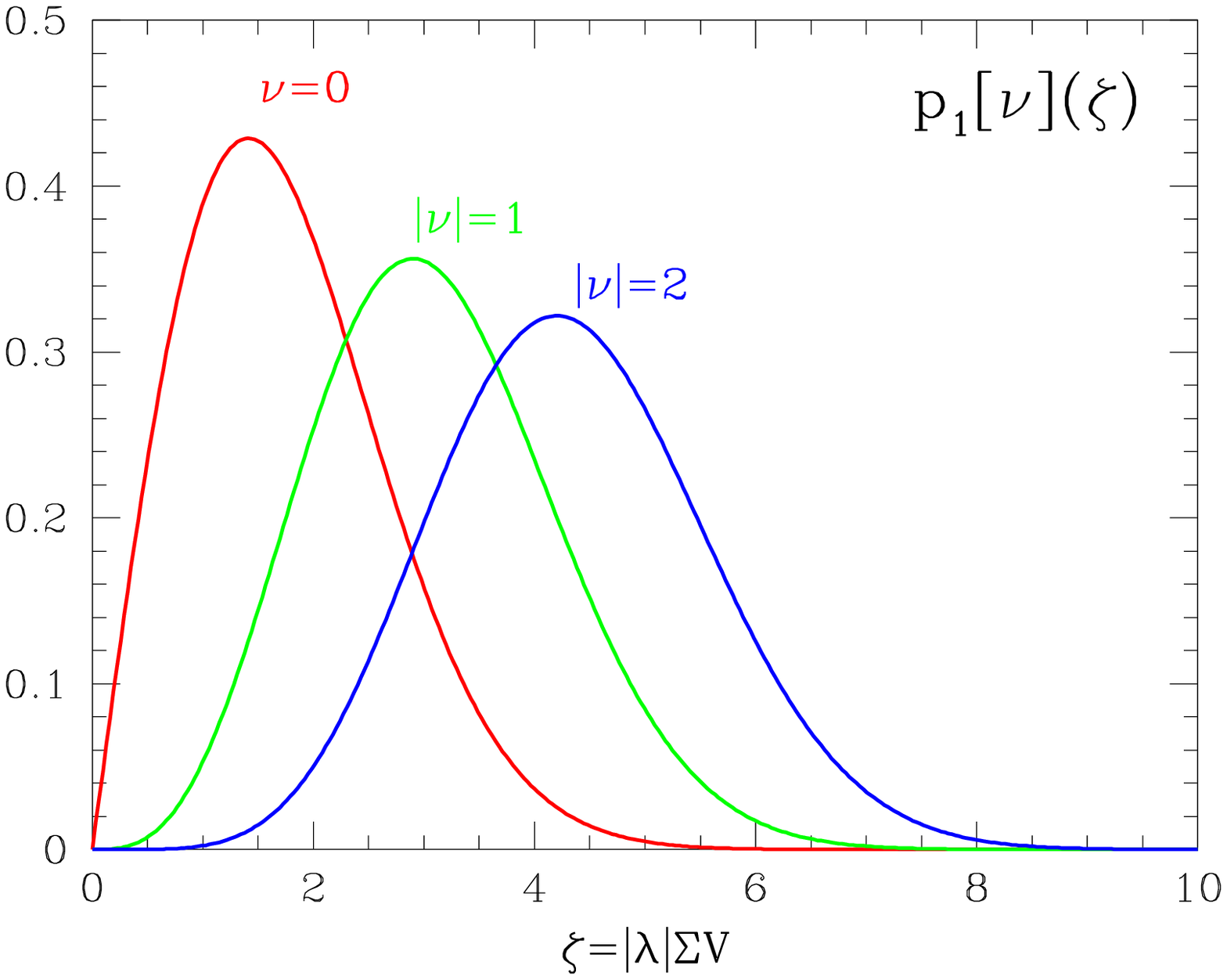,height=6.0cm,width=6.5cm,angle=0}} \\
\end{tabular}
\caption{
Probability density distributions of low-lying eigenvalues 
in quenched QCD. Left: distributions of the first four 
eigenvalues ${\rm p}_i(\zeta)$ $(i=1,\dots 4)$ and of the 
microscopic total density $\rho_s(\zeta)$ for $\nu=0$. 
Right: distributions of the first eigenvalue for 
$|\nu|=0,1,2$.\label{fig:rmt}}
\end{center}
\end{figure}

\begin{figure}[thb]
\vskip 0.8cm 
\begin{center}
\mbox{\epsfig{file=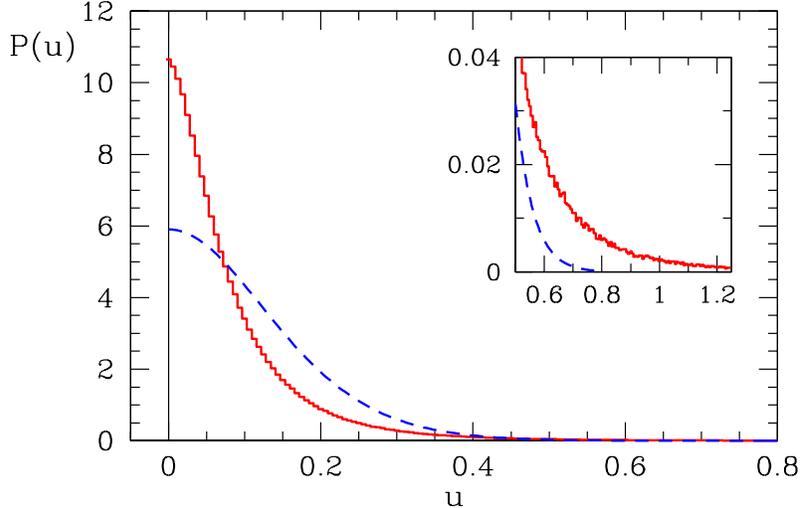,height=7.0cm,width=11cm,angle=0}} \\
\caption{Probability distribution of $u=|{\rm Re}[v_{11}(x)]| V$ (cf.~Eq.(\ref{eq:vkl})) 
for the first norma\-lized non-zero mode eigenvector of $D$. 
The solid line is the distribution
obtained at $\beta=6.0$ on a $16^4$ lattice (Lattice B, see below), 
while the dashed curve is the one 
expected for a random vector model.\label{fig:wavefunc}}
\end{center}
\end{figure}

Although the technique we are going to describe can be more 
widely applied, we restrict ourselves to lattices of 
spacing $a$, volume $V=T L^3$ and with periodic boundary 
conditions imposed on all fields.
QCD gluons and fermions are discretized using the standard plaquette Wilson 
action and the Neuberger-Dirac operator $D$, 
respectively. The latter \cite{Neuberger} satisfies the 
Ginsparg-Wilson relation \cite{Ginsparg:1982bj}
\be\label{eq:GW}
   \gamma_5 D+D \gamma_5=\bar{a} D \gamma_5 D\, , 
\ee
and thus preserves an exact chiral symmetry at 
finite lattice spacing \cite{Luscher:1998pq}. The Neuberger 
operator, the parameter 
$\bar{a}$ and other conventions used 
in this section are defined in Appendix A, and are the 
same as in Ref.~\cite{Giusti:2002sm}. The massive
lattice Dirac operator is given by
\be
  D_m=(1-\frac{\bar{a}m}{2})D+m
\ee
where $0\leq\bar{a} m\leq2$.
For a given gauge configuration the massless Dirac 
operator can be diagonalized, and chirality 
implies that non-chiral modes appear in complex 
conjugate pairs, i.e.
\ba
D \eta_{\lambda_k} & = & \lambda_k\, \eta_{\lambda_k}\, , \qquad k=1,2,\dots\, , \\
D \eta_{\lambda^*_k} & = & \lambda^*_k \eta_{\lambda^*_k}\, , \qquad 
\eta_{\lambda^*_k} = \gamma_5 \eta_{\lambda_k}\, .
\ea
Random matrix theory \cite{Shuryak:1992pi,Nishigaki:1998is} 
(and to some extent quenched ChPT \cite{Akemann:2003tv}) 
predicts the probability distribution of each eigenvalue in 
the low-lying end of the spectrum to be a function of the rescaled 
variable $\zeta = |\lambda| \Sigma V$ only. 
Some of these predictions are reported in 
Fig.~\ref{fig:rmt} for the quenched theory. In particular,
the predicted individual distributions of 
the first four eigenvalues together with the total microscopic 
density for $\nu=0$ are shown in the plot on the left, while 
the individual distributions of the first 
non-zero eigenvalue are shown for several values 
of $\nu$ in the right plot. 
It is interesting to note that there is no gap for small values 
of $\zeta$, and the total distribution is predicted to be 
$\rho_s(\zeta)\sim \zeta^{2(|\nu|+N_{\rm f})+1}$. Therefore
arbitrarily small eigenvalues can occur (either in the full or the quenched theory) 
with a probability which decreases exponentially with $|\nu|$ and $\Nf$. 
The expectation value of the lowest eigenvalue and the level 
splittings near the origin are of $O(1/\Sigma V)$, as can be seen in 
Fig.~\ref{fig:rmt}. The ratios of expectation values of low-lying
eigenvalues are then parameter-free predictions of RMT, and they have been
confronted with quenched QCD data in Ref.~\cite{Giusti:2003gf}. 
An impressive agreement has been found for volumes larger than 
about $5$~fm$^4$.\\ 
\indent No analytic 
predictions are available so far for the properties 
of the corresponding wave functions. In 
Ref.~\cite{Giusti:2003gf} it was found that the probability 
distribution of their norm at a fixed lattice site is 
broader than the one expected for a normalized random 
vector\footnote{The behaviour of the eigenfunctions corresponding 
to the low modes of the Dirac operator has already been studied numerically in 
different contexts, see Ref.~\cite{Edwards:2001ei}.}. 
Since in the following we are interested in the correlation
function of two left-handed currents, we have studied the probability  
distribution of $u=|{\rm Re}[v_{11}(x)]| V$, where
\be\label{eq:vkl}
v_{kl}(x) = [\eta_{\lambda_k}^\dagger \gamma_0 P_- 
            \eta_{\lambda_l}](x) \, ,
\ee
on a lattice of volume $16^4$ at $\beta=6.0$ 
(Lattice B, see next section).
The result is shown in Fig.~\ref{fig:wavefunc} together with the 
analogous prediction for the case that the $\eta$'s are 
treated as random vectors with unit norm and such that 
$\sum_x [\eta_{\lambda_k}^\dagger \gamma_5 \eta_{\lambda_k}](x)=0$.
As can be seen from the inset 
in the figure, the distribution decreases
quite slowly, and thus the probability of finding a point on the lattice
for which $u$ exceeds its mean value by far is quite high: for instance,
the probability for finding $u\geq0.5$ is $0.7\%$. We also mention that 
the distribution of $|{\rm Im}[v_{11}(x)]| V$ nicely overlaps with the one 
reported in Fig.~\ref{fig:wavefunc}. \\ 
\indent These properties of eigenvalues and eigenfunctions
of the Dirac operator imply that, when a fermionic correlation function 
is computed with standard Monte Carlo techniques, the relative contributions from the 
various eigenvalues can change dramatically with $m$:
\begin{itemize}
\item for $m\gg 1/\Sigma V$, the low-lying end of the spectrum of $D_m$ 
is very dense near $m$, and (many) contributions from the corresponding wave-functions 
are averaged with essentially the same weight.
\item for $m \sim 1/\Sigma V$, the mass is comparable with the expectation value
of the modulus of the lowest eigenvalue of $D$ (see Fig.~\ref{fig:rmt}), and therefore 
the low-lying spectrum of $D_m$ appears discrete (the splittings of the 
eigenvalues are of the same order as their values). The
contribution of just a few eigenvectors to a given observable can be 
substantial and their space-time fluctuations can then induce large fluctuations 
in its estimate. A significant improvement can be achieved in this 
case by low-mode averaging as we explain in the remainder of this section.
\item  for $m\ll 1/\Sigma V$, the mass is much smaller than the expectation value 
of the lowest eigenvalue of $D$. Extremely small eigenvalues of $D_m$ can then 
occur with a small but non-negligible probability. The standard Monte Carlo 
sampling of the path integral for fermionic correlation functions is 
problematic in this case, because a sizeable contribution to the estimate 
of these integrals is obtained from configurations that have a small statistical 
weight (i.e., those with the smallest eigenvalues of $D_m$, for which these
observables are very large). We did not find a straightforward solution 
to this problem. It would probably require an improved 
algorithm where the low-mode contribution to the observable is included in 
the Monte Carlo sampling of the integral. If such an algorithm exists, 
it will presumably be very expensive in practice. 
\end{itemize}

The information above can be exploited in order to develop 
efficient algorithms for computing fermionic correlation 
functions in the regime $m\gsim 1/\Sigma V$. In the following 
we focus on the two-point function 
\be\label{eq:jljl}
C^{ab}(t) = \sum_{\bf \vec x} \langle J^a_0(x) J^b_0(0) \rangle 
\ee
of two left-handed charge densities, 
\be\label{eq:j0qcd}
J^a_0(x) = \bar\psi(x) \mbox{T}^a \gamma_0 P_- \tilde \psi(x)\; ,
\ee
where $\tilde \psi$ is defined in Appendix A. Writing 
$C^{ab}(t)=\Tr [\mbox{T}^a \mbox{T}^b] C(t)$, and using the spectral 
decomposition, we get
\ba
C(t) & = & -\sum_{\bf \vec x}  \Big\langle \sum_{k,l} 
\frac{(\lambda_k-\lambda^*_k) (\lambda_l-\lambda^*_l)}
{|\bar\lambda_k|^2|\bar\lambda_l|^2}
[\eta_{\lambda_k}^\dagger \gamma_0 P_- \eta_{\lambda_l}](x) 
[\eta_{\lambda_l}^\dagger \gamma_0 P_- \eta_{\lambda_k}](0) 
\Big\rangle\label{eq:first_dec} \\
 & = & - \sum_{k,l} \sum_{\bf \vec x} \Big\langle 
\frac{(\lambda_k-\lambda^*_k) (\lambda_l-\lambda^*_l)}
{|\bar\lambda_k|^2|\bar\lambda_l|^2}
[\eta_{\lambda_k}^\dagger \gamma_0 P_- \eta_{\lambda_l}](x) 
[\eta_{\lambda_l}^\dagger \gamma_0 P_- \eta_{\lambda_k}](0) 
\Big\rangle\, ,\label{eq:secon_dec}
\ea
where $\bar \lambda_k =\{(1-\bar a m/2) \lambda_k + m\}$
are the eigenvalues of the massive operator $D_m$.\\
\indent The standard way of estimating $C(t)$, by computing 
for every gauge configuration the propagator from a local source 
to any other point, turns out to be efficient in the 
mass range $m\gg 1/\Sigma V$, where the correlator is the 
result of an average over many comparable contributions from 
different eigenfunctions. When the quark mass reaches the 
$\epsilon$-regime, i.e.  $m\sim 1/\Sigma V$, the presence of 
``bumps'' in single wave functions, associated with 
low-lying eigenvalues, can generate much larger statistical fluctuations.
They can be reduced by noting that {\it each contribution} to the sum 
over $k$ and $l$ in Eq.~(\ref{eq:secon_dec}) {\it can be estimated independently}. 
The statistical error of each term can be reduced by increasing the number of 
configurations, exploiting the symmetries of the theory, etc.\\
\indent In particular, 
the variance of the low-mode contribution to $C(t)$ 
can be decreased by exploiting the translational invariance of 
the theory. It is important to note that, based on the techniques developed in 
Ref.~\cite{Giusti:2002sm}, this can 
be achieved {\it without} computing 
low-lying eigenmodes and eigenfunctions with {\it very high precision}, 
which can be numerically expensive, especially with Neuberger fermions.\\
\indent We now describe a particular implementation of the ideas sketched above,
which we have adopted for computing $C(t)$.
For each gauge configuration we have computed the topology $\nu$, and,
starting from a set of Gaussian random sources, we have minimized 
the Ritz functional until the estimated relative 
errors of the calculated eigenvalues drop below a specified bound 
$\omega_k$ \cite{Giusti:2002sm}, cf.~Eq.(\ref{eq:ritz}). This guarantees that
the orthonormal Dirac fields $u_1\cdots u_n$, 
which approximate the eigenvectors, satisfy
\be\label{eq:ritz}
A u_k = \alpha_k u_k + r_k\, , \qquad \left(u_l,r_k\right)=0\, , \qquad 
\|r_k\|\leq\omega_k \alpha_k\, ,
\ee
for all $k,l$. In Eq.~(\ref{eq:ritz}), $A=P_s D_m^\dagger D_m P_s$,
$P_s$ ($s=\pm$) is the projector into the chirality sector without zero modes, 
and $\alpha_k$ are the approximate eigenvalues of $A$.
We can then define a ``subtracted'' left-left propagator $P_- {\cal S}^h (x,y)P_+$ for 
the massive Neuberger operator (which we have computed in practice as described 
in Ref.~\cite{Giusti:2002sm}) through the equation 
\be
P_- {\cal S}(x,y) P_+ = P_- \Big[ \sum_{k=1}^{n} \frac{1}{\alpha_k} 
e_k(x) e_k(y)^\dagger   +  {\cal S}^h(x,y) \Big] P_+
\ee
where
\be
{\cal S}(x,y) = \frac{1}{1-\bar a m/2}\{ D^{-1}_m\}(x,y)
\ee
and\footnote{Unlike $\eta_{\lambda_k}$, the two chiral components of 
$e_k$ are not normalized, in order to simplify our formul\ae.} 
\be
e_k = P_s u_k + P_{-s} D P_{s} u_k\, .
\ee
The two-point correlation function in Eq.~(\ref{eq:jljl}) is then given by
\be\label{eq:split}
C(t) = C^{ll}(t) + C^{hl}(t) + C^{hh}(t)\, ,
\ee
where
\ba\label{eq:split1}
C^{ll}(t) & = &  - \sum_{k,l=1}^{n} \sum_{\bf \vec x} 
\Big\langle 
\frac{ [e^\dagger_k\gamma_0 P_- e_l](x) [e^\dagger_l\gamma_0 P_- e_k](0) }
{\alpha_k\alpha_l}\Big\rangle\, ,
\\
C^{hl}(t) & = & - \sum_{k=1}^{n} \sum_{\bf \vec x}
 \Big\langle \frac{1}{\alpha_k} e_k^\dagger(x) \gamma_0 P_- {\cal S}^h(x,0) 
\gamma_0 P_- e_k(0) \Big\rangle +  (x \leftrightarrow 0)\, ,\label{eq:split2}\\
C^{hh}(t) & = & -\sum_{\bf \vec x}\Big\langle \mbox{Tr} 
\left[\gamma_0 P_-{\cal S}^h(x,0)\gamma_0 P_- {\cal S}^h(0,x)\right] \Big\rangle\; ,
\label{eq:split3}
\ea
and the space-time dependences of eigenvectors and propagators have been 
shown explicitly. By noticing that the starting vectors of the 
Ritz minimization procedure are extracted with a 
translationally invariant action, it is straightforward to prove that each 
contribution on 
the right-hand side of Eqs.~(\ref{eq:split1})-(\ref{eq:split3}) is 
translationally invariant 
even if the vectors $u_k$ in Eq.~(\ref{eq:ritz}) are only approximate 
eigenvectors of $A$, i.e. $\omega_k\neq 0$. In this case, 
in addition to the gluon field, 
also the random vectors needed to start the Ritz functional minimization should
be 
translated\footnote{We thank M.~L\"uscher for having 
clarified this point to us.}. 
Therefore, 
\ba
C^{ll}(t) & = &  -\frac{1}{V} \sum_{k,l=1}^{n} 
\sum_{x,y} \delta_{t,t_x-t_y}\,
\Big\langle 
\frac{ [e^\dagger_k\gamma_0 P_- e_l](x) [e^\dagger_l\gamma_0 P_- e_k](y) }
{\alpha_k \alpha_l}\Big\rangle  \, ,
\\
C^{hl}(t) & = & - \frac{1}{L^3} \sum_{k=1}^{n} \sum_{x, {\bf \vec y}}
\delta_{t,t_x-t_y} 
\Big\langle \frac{1}{\alpha_k} e_k^\dagger(x) \gamma_0 P_- {\cal S}^h(x,y) 
\gamma_0 P_- e_k(y)\Big\rangle + (x \leftrightarrow y)\label{eq:hl}
\ea
hold independently of the number $n$ of eigenvectors which have been subtracted 
and of the precision $\omega_k$ they have been determined with. By contrast, the statistical 
variance of the signal changes with $n$ and $\omega_k$. Note that the computation of 
$C^{hl}(t)$ can be quite expensive since it requires an inversion of the Dirac operator
for every vector $e_k$.

\section{Numerical results}\label{sec:numeric}
To test the procedure described in the previous section, we have simulated two lattices 
at $\beta=6.0$ ($a\simeq 0.09$~fm) with volumes
$V=24 \times 16^3$ (A) and $V=16^4$ (B). Since the generation of 
gauge-field configurations consumes a negligible amount of computer 
time, we have performed many update cycles between subsequent measurements 
so that they can be assumed to be statistically independent. The computation 
of the index, the low-lying eigenvalues and the inversion of the 
Neuberger-Dirac operator have been carried out using the techniques reported in 
Ref.~\cite{Giusti:2002sm}. For each gauge-field configuration of the lattice A(B), 
7(6) low-lying eigenvalues of $D^\dagger D$ have been extracted in the chirality 
sector without zero modes with a relative uncertainty of $\omega_k=0.05$ (see below), 
while a further one has been determined with lower precision to stabilize the 
Ritz functional minimization. The subtracted and the full propagator have 
been computed by requiring a residue of $5\cdot 10^{-7}$
in the adaptive conjugate gradient.\\
\indent Lattice A has been devoted to studying the left-left 
correlation function in the $p$-regime: 
we have generated 113 gauge configurations, 
inverted $D_m$ for masses $am=0.025,0.040,0.060,0.080,0.100$, and 
computed the correlation functions in the standard manner 
(local source and sink) and with the 
low-mode averaging (LMA) procedure described in the previous section. After symmetrizing the 
correlators around $t=T/2$, we estimated the statistical errors by a jackknife procedure
and fitted the correlation functions with the expression given in Eq.~(\ref{eq:Ct_p})
in the time interval $6-12$. The lower limit was fixed at the point where we 
found stabilization of the effective meson masses. The results of the fits are given 
in Table~\ref{tab:p-results}. They 
are compatible with previous computations \cite{Giusti:2001pk,Hernandez:2001hq}
(for recent reviews see \cite{Hernandez:2001yd,Giusti:2002rx}) within the 
statistical uncertainties. In this regime and at this volume, 
the benefit of the low-mode averaging is visible for the two
lightest quark masses only, and it is likely to be 
less effective when the volume increases and the quark mass is kept fixed.\\
\begin{table}[thb]
\vskip 0.5cm
\begin{center}
\begin{tabular}{c|cc|cc}
\hline
  &  \multicolumn{2}{c|}{LMA} & \multicolumn{2}{c}{local}\\
\hline
   $a m$   & $a M^V_P$ &    $a F^V_P$   & $a M^V_P$ &    $a F^V_P$ \\
\hline
   0.025   & 0.199(6)  &  0.0341(6) & 0.198(8) &  0.0336(9) \\
   0.040   & 0.242(5)  &  0.0355(6) & 0.244(7) &  0.0349(7) \\
   0.060   & 0.292(5)  &  0.0374(6) & 0.295(6) &  0.0369(6) \\
   0.080   & 0.335(4)  &  0.0392(6) & 0.339(5) &  0.0390(5) \\
   0.100   & 0.375(4)  &  0.0410(6) & 0.380(5) &  0.0410(5) \\
\hline
\end{tabular}
\caption{Meson masses and decay constants from 
the left-left correlators computed in the $p$-regime with 
the LMA procedure, and with the standard local one. By fixing the lattice 
spacing with $r_0$ from Ref.~\cite{Necco:2001xg},
the physical value of the Kaon mass $M_K=496$~MeV, 
expressed in lattice units, would correspond to 
$a M_K\approx 0.234$.
\label{tab:p-results}}
\end{center}
\end{table}
\begin{figure}[thb]
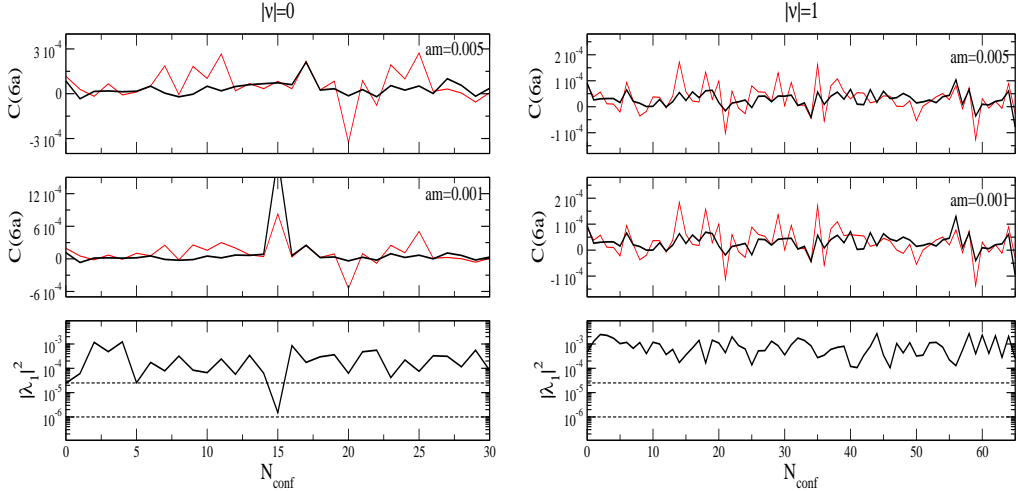

\vskip 1.0 cm
\begin{center}
\begin{tabular}{cc}
\mbox{\epsfig{file=jljl_t6_nu0.eps,width=6.5cm,height=6.5cm,angle=0}} &
\mbox{\epsfig{file=jljl_t6_nu1.eps,width=6.5cm,height=6.5cm,angle=0}} \\
\end{tabular}
\caption{Monte Carlo histories (Lattice B) for the absolute value squared of the 
lowest non-zero eigenvalue of $D$ (bottom) and for the left-left correlators 
computed at $t/a=6$: $|\nu|=0$ (left) and $|\nu|=1$ (right), $a m=0.001$ (middle)
and $a m=0.005$ (top). The dashed lines in the eigenvalue plots 
indicate the two values of $(a m)^2$. In the plots of the 
correlation functions, the thick lines are obtained with LMA, while the 
thin ones in the standard manner.\label{fig:jljl_t6_eps}}
\end{center}
\end{figure}
\indent The values of $F_P$ follow a remarkable linear 
behaviour in the quark mass. A linear fit of the form 
$a F_P = A_1 + A_2 \cdot (a m)$ gives
\ba
A_1 =  0.0318(7) \;\, \, ,  & & A_2 =0.093(6) \qquad \,\; \mbox{LMA}\\
A_1 =  0.0310(10)\, , & & A_2 =0.100(10) \qquad \mbox{Local}\; .
\ea
Quadratic fits to the data give results very well compatible with the previous ones, 
with the coefficients of the quadratic terms compatible with zero.\\
\indent Lattice B has been reserved for the $\varepsilon$-regime: we have 
generated $203$ independent gauge configurations of which 31, 66 and 
44 have topological charge $|\nu|=0,\, 1,\, 2$, respectively. In these topological sectors we  
have computed the quark propagators for masses $a m=0.001,0.005,0.010$ 
which correspond to $m \Sigma V \approx 0.11,\, 0.55,\, 1.1$ respectively, if 
the bare quark condensate $\Sigma$ is taken from the analogous lattice of 
Ref.~\cite{Giusti:2003gf}. As in the previous case, the two-point 
correlators of the left-handed current have been computed 
in the standard manner and with LMA.\\
\indent In Fig.~\ref{fig:jljl_t6_eps} the Monte Carlo histories for the 
left-left correlator at $t/a=6$ and for the lowest non-zero eigenvalue 
of $D^\dagger D$ are reported
for $|\nu|=0,1$ and $am=0.005,\, 0.001$. For the lightest mass, a spike 
in $C(t)$ is clearly visible in correspondence with a very low (the lowest
produced) eigenvalue, which happens to be roughly one order of magnitude 
lower than its expectation value \cite{Giusti:2003gf}. A closer look at this 
configuration reveals that the spiky contribution is indeed due to the light-light
contribution $C^{ll}(t)$ and is not cured with the LMA procedure proposed here. 
\begin{figure}[thb]
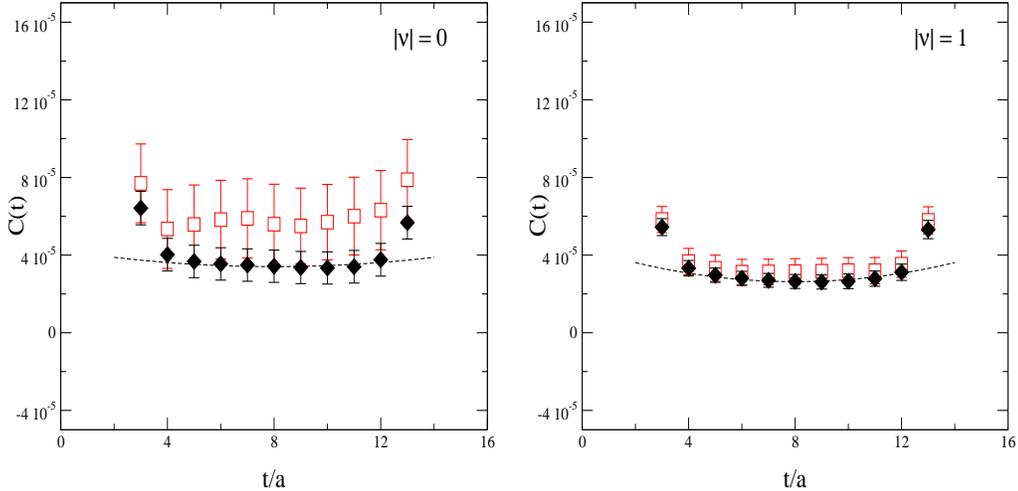

\vskip 1.0 cm
\begin{center}
\begin{tabular}{cc}
\mbox{\epsfig{file=jljl_nu0.eps,width=6.5cm,height=6.5cm,angle=0}} &
\mbox{\epsfig{file=jljl_nu1.eps,width=6.5cm,height=6.5cm,angle=0}} \\
\end{tabular}
\caption{Left-left correlators in the $\epsilon$-regime for $am=0.005$,
$|\nu|=0$ (left) and $|\nu|=1$ (right), computed with LMA (filled diamonds) and in the 
standard manner (open squares). The dashed lines represent
fits to LMA data (see main text).
\label{fig:jljl_eps}}
\end{center}
\end{figure}
As expected (cf.~Sect.~\ref{sec:lowm}), for $m\ll 1/\Sigma V$ the Monte Carlo history 
shows evidence for extreme statistical fluctuations, 
and therefore we discard data at the lightest mass in the 
following analysis. The spiky behaviour disappears for the two heavier masses 
and for them the Monte Carlo history is well-behaved.\\
\indent For masses $m\sim 1/\Sigma V$, the LMA estimate of 
the correlation function $C(t)$ is indeed less fluctuating than the one 
computed in the standard manner.
Its variance turns out to be roughly a factor 
two smaller for the topologies and masses that we consider, 
as shown in Fig.~\ref{fig:jljl_eps} for $|\nu|=0,1$ and $am=0.005$. 
The variance reduction is a function of 
the number of eigenvalues extracted $n$, and the 
precision $\omega_k$ they have been computed with. It can also vary 
depending on which contributions are chosen to be averaged over.
We tried several values of $n$ and $\omega_k$ and the ones used in this 
computation turn out to be a good compromise between the gain in 
statistics and the additional computational cost (roughly a factor 2), which
is mainly due to the computation of the mixed contribution $C^{hl}(t)$. 
A more systematic optimization of these parameters is desirable but goes 
beyond the scope of this paper.\\
\indent We fitted the LMA correlations with the expression given in
Eq.~(\ref{eq:eps_quenched}),
\be\label{eq:eps_fit}
C(t) = \frac{B_1^2}{2 T}\,\Big[ 1 + B_2\, h_1\left(\frac{t}{T}\right) \Big]\, ,
\ee
for each topological sector and mass.
The lower temporal limit was fixed at $t/a=5$, a point where we 
found stabilization of the $\chi^2$ for all correlators. 
The results obtained and the associated 
errors computed with a jackknife procedure are reported 
in Table \ref{tab:eps-results}. 
\begin{figure}[htb]
\begin{center}
\mbox{\epsfig{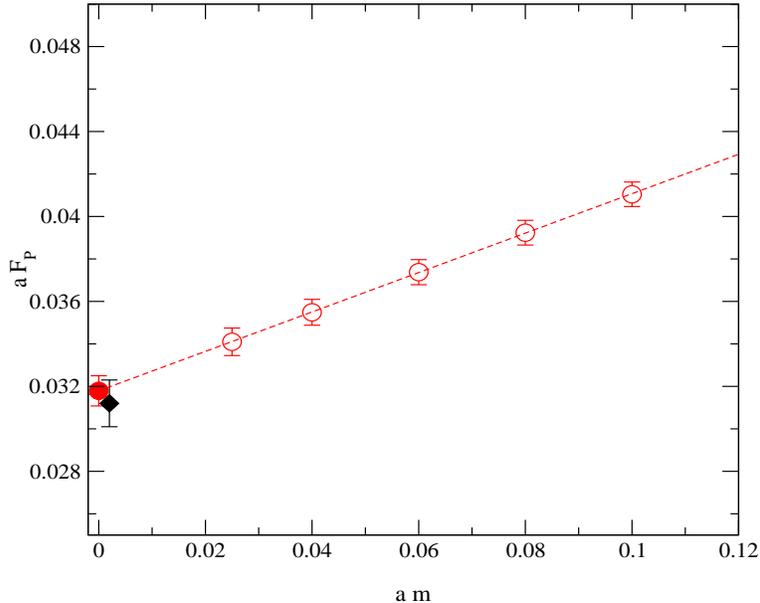}} \\
\caption{Results obtained in the $p$-regime for $F_P$
(open circles) and in the $\epsilon$-regime for $F$ (filled diamond). The dashed 
line represents a linear fit to the $p$-regime points. The corresponding 
intercept is also shown (filled circle).
\label{fig:Fp_m}}
\end{center}
\end{figure}
\begin{table}[thb]
\begin{center}
\begin{tabular}{ccccccc}
\hline
   $a m$   & $|\nu|$ & &   $B_1$  & &  $B_2$ & \\
\hline
           &    0    & &0.033(3)  & & 3(2) & \\
  0.010    &    1    & &0.032(2)  & & 5(1) & \\
           &    2    & &0.030(2)  & & 5(1) & \\
\hline
           &    0    & &0.034(4)  & & 2(2) & \\
  0.005    &    1    & &0.032(2)  & & 4(1) & \\
           &    2    & &0.030(2)  & & 4(1) & \\
\hline
\end{tabular}
\caption{Results from the $\epsilon$-regime data of a fit of the form given 
in Eq.~(\ref{eq:eps_fit}).\label{tab:eps-results}}
\end{center}
\end{table}
Within the large statistical errors,
the values for $B_2$ are compatible with the 
prediction of Eq.~(\ref{eq:eps_quenched}) and the estimate of
the condensate extracted in Ref.~\cite{Giusti:2003gf}.
The results for $B_1$ are very well compatible for all masses 
and topological sectors. We estimate our best value of the decay 
constant in the $\epsilon$-regime, $aF=0.0312(11)$, by averaging the two values 
of $B_1$ in each topological sector and then averaging the 
results as independent determinations. This value is  
in very good agreement with the linearly extrapolated value  
from the $p$-regime, as shown in Fig.~\ref{fig:Fp_m}, and constitutes
one of the main results of this paper.\\
\indent By renormalizing the local left-handed current with $Z_J=1.55$ \cite{Giusti:2001pk}
and by fixing the lattice spacing with $r_0$ from Ref.~\cite{Necco:2001xg},
we obtain  $F=104(2)$~MeV and $F=102(4)$~MeV in the $p$ and $\epsilon$-regimes, 
respectively. These values are compatible within 2$\sigma$ yet more precise
 than the one we found recently in Ref.~\cite{Giusti:2003iq} from a
 study of the topological zero-mode wave functions. Note however that
 the systematic error due to the finite lattice spacing has not been
 quantified in the present study. The agreement between these two
 determinations would provide a further check of our main working assumption that
 quenched ChPT describes the low-energy regime of quenched QCD in certain
 ranges of kinematical scales. 
By fitting $F_P$ linearly as a function of $(M^V_P)^2$, we also obtain our 
estimate of the LEC in Eq.~(\ref{eq:FPq}) to be $\alpha_5=1.66(8)$. This value is in 
the same ballpark as the one found in Ref.~\cite{Bardeen:2003qz}, while it is higher 
than the one obtained in Ref.~\cite{Heitger:2000ay}. 
To understand the discrepancy we have analyzed our data as suggested
in Ref.~\cite{Heitger:2000ay}, and we have found $\alpha_5=1.08(5)$,
which agrees well with their value of $0.99(6)$. Closer inspection
shows that the difference can be traced back to two features of the
analysis presented in~\cite{Heitger:2000ay}: the first is the Taylor
expansion of $1/(1+\alpha_5 y_{\mbox{\scriptsize ref}}/2)$ to leading
order in $\alpha_5$. Although such an expansion is justified, since
the difference with the exact result is a higher-order effect, 
the resulting ambiguity is large and produces an estimate
for $\alpha_5$ which is smaller by
20--25\%. Indeed, this discrepancy was included as a systematic
uncertainty in~\cite{Heitger:2000ay}. The remaining difference can be
explained by the fact that in Ref.~\cite{Heitger:2000ay} the value $F$
was constrained to coincide with the experimental value of the pion
decay constant, which is smaller by about 15\% than typical
quenched estimates. This assumption is not in accord with 
our working hypothesis that quenched data should be 
reproduced by quenched ChPT.\\
\indent The errors quoted above for the LECs include only our
statistical errors. A more detailed assessment of the various
systematic errors would require computations at different volumes
and lattice spacings, which goes beyond the scope and primary
goal of this study.

\section{Conclusions}\label{sec:concl}
The low-mode averaging technique proposed in this paper reduces
large statistical fluctuations in correlation functions 
due to the presence of local ``bumps'' in the wave functions 
associated with the low-lying eigenmodes of the Dirac operator. 
When applied to the two-point function of the left-handed vector
current in the region of quark masses $m\sim 1/\Sigma V$, 
it provides an estimate of the correlator with a variance significantly 
reduced with respect to the standard one. As a result
the $\epsilon$-regime of QCD can be safely reached in all 
topological sectors.\\
\indent It is conceivable that more involved correlation functions such as
singlet diagrams (see for example \cite{Neff:2001zr}) or those needed for non-leptonic weak decays 
can also benefit from low-mode averaging. Conventional formulations 
of lattice QCD may profit as well from this technique. For instance, it could mitigate the 
problem of exceptional configurations encountered for Wilson fermions, or speed up unquenched
simulations.\\
\indent By matching the quenched QCD results for the left-left correlators 
computed for quark masses in the $p$- and $\epsilon$-regimes with those 
of quenched ChPT, we  
obtained estimates for the quenched low-energy constants $F$ and $\alpha_5$.
The agreement we found between the value of the pseudoscalar decay constant
extrapolated from the $p$-regime and the one extracted directly in 
the $\epsilon$-regime is remarkable.\\
\indent In Ref.~\cite{Giusti:2003iq}, we studied the possibility of using the contribution
of topological zero-mode wave functions to the pseudoscalar correlator, 
in order to extract $F$. As far as the convergence of the chiral expansion is 
concerned, it was found that in full QCD (assuming $F \approx 93$~MeV), 
one would need to go to a lattice extent $\gsim 2.0$~fm ($\gsim 2.5$~fm)
to have the first non-trivial relative correction to be less than 
50\% (30\%). Inspecting the constant part of the $\epsilon$-regime 
expression in~Eq.~(\ref{eq:jljl-e}), we find that for the observable studied 
in the present paper, the same relative corrections would be obtained
already with lattice extents $\gsim 1.4$~fm ($\gsim 1.8$~fm). 
Therefore, 
current correlators near the chiral limit should allow for smaller
systematic uncertainties in the extraction of the LEC $F$ at realistically 
accessible volumes than the zero-mode wave functions. It would be interesting 
to study whether the same remains true for other LECs as well, 
such as those related to weak decays.\\
\indent In the quenched theory, on the other hand, systematic errors are very 
hard to quantify, but the apparent convergence of our expression for the 
current correlator compares well with the one 
observed for the zero-mode wave functions in Ref.~\cite{Giusti:2003iq}, 
while having at the same time the advantage that the singlet 
parameters $m_0^2/2N_c$, $\alpha/2N_c$ of quenched chiral perturbation 
theory do not enter at all at this order.\\ 

\section*{Acknowledgements}
The present paper is part of an ongoing project whose final goal
is to extract low-energy parameters of QCD from numerical simulations 
with GW fermions. The basic ideas of our approach were developed 
in collaboration with M. L\"uscher; we would like to thank him for
his input and for many illuminating discussions. 
We are also indebted to P.H.\ Damgaard, C. Hoelbling, K.~Jansen and 
L.\ Lellouch for interesting discussions.\\ 
\indent The simulations were performed on PC clusters 
at the  Leibniz-Rechenzentrum der Bayerischen Akademie der 
Wissenschaften, the 
Max-Planck-Institut f\"ur Physik in Munich,
the Max-Planck-Institut f\"ur Plasmaphysik in Garching, 
and at the Valencia University.
We wish to thank all these institutions for supporting our project
and the staff of their computer centers for technical help.
L.~G.~was supported in part by the EU under contract 
HPRN-CT-2002-00311 (EURIDICE), and 
P.~H.\ by the CICYT (Project No.\ FPA2002-00612) and 
by the Generalitat Valenciana (Project No.\ CTIDIA/2002/5).  

\appendix 

\section*{Appendix A. Some definitions for the GW fermions}
In this paper we employ the same conventions as 
in Ref.~\cite{Giusti:2002sm}. The Dirac matrices satisfy
\be
  (\gamma_{\mu})^{\dagger}=\gamma_{\mu},\qquad
  \{\gamma_{\mu},\gamma_{\nu}\}=2\delta_{\mu\nu},
\ee
and we have chosen a chiral representation with
\be
  \gamma_{5}=\gamma_{0}\gamma_{1}\gamma_{2}\gamma_{3}
  =\pmatrix{1&0\cr 0&-1\cr}.
\ee
The chiral projectors are defined as
\be
P_{\pm}=\frac{1}{2}(1\pm \gamma_5)\, .
\ee
The Wilson-Dirac operator is given by
\be
  D_{\rm w}=\frac{1}{2}\left\{
  \gamma_{\mu}(\nabla^*_{\mu}+\nabla_{\mu})-a\nabla^*_{\mu}\nabla_{\mu}\right\},
\ee
where
\ba
  \nabla_{\mu}\psi(x) & = & \frac{1}{a}
  \left\{U(x,\mu)\psi(x+a\hat{\mu})-\psi(x)\right\}\, , \\
  \nabla^*_{\mu}\psi(x)& = & \frac{1}{a}
  \left\{\psi(x)-U(x-a\hat{\mu},\mu)^{-1}\psi(x-a\hat{\mu})\right\}
\ea
are the gauge-covariant forward and backward difference
operators, $a$ denotes the lattice spacing,
$U(x,\mu)\in {\rm SU(3)}$ are the link variables and
$\hat{\mu}$ is the unit vector along the direction $\mu$.
The Neuberger-Dirac operator is defined as \cite{Neuberger}
\ba
D & = & \frac{1}{\bar a}\left\{1+\gamma_{5}\,\mbox{sign}(Q)\right\},\\
Q & = & \gamma_{5}\left(a D_{\rm w} -1-s\right), \qquad
  \bar a =\frac{a}{1+s},
\ea
where $s$ is a real parameter in the range $|s|<1$.
It satisfies the Ginsparg-Wilson relation in Eq.~(\ref{eq:GW}).
Infinitesimal chiral transformations of the 
fermion field are given by \cite{Luscher:1998pq}
\be
\delta \psi = \gamma_5 (1-\bar a D)\psi\, , \qquad
\delta \bar \psi =\bar \psi \gamma_5 \; .
\ee
The modified fermion field 
\be
\tilde \psi = (1-\frac{1}{2}\bar a D)\psi
\ee
transforms according to 
\be
\delta \tilde \psi = \gamma_5 \tilde \psi\, ,
\ee
and therefore if a composite operator
is defined using $\tilde\psi$ instead of $\psi$, it has 
the same transformation behaviour as the corresponding one in the 
continuum. 

\section*{Appendix B. Pseudoscalar mass in the $p$-regime of ChPT to NLO}
For completeness we report in this appendix NLO expressions for 
the pseudoscalar meson mass in the $p$-regime of ChPT. We again
consider the case of degenerate light quarks only.
The effective finite-volume pion mass $M_P^V$, entering the 
prediction for the correlation 
function of two left currents reported in Eq.~(\ref{eq:Ct_p}),
is given by
\be
M_P^V  \equiv  M_P \biggl( 
 1 + \frac{g_1}{2 \Nf F_P^2} \biggr) 
 \;, \label{eq:MPV}
\ee
where at the same order the infinite-volume mass $M_P$ is
\be
M_P =  M 
 \biggl[
 1 +  \frac{G(M^2)}{2 \Nf F^2} - \frac{M^2}{2 (4\pi F)^2}
 \Bigl( \Nf \alpha_4 + \alpha_5 - 2 \Nf \alpha_6 - 2 \alpha_8 \Bigr)
 \biggr]\; .
\ee
The function $g_1$ reads~\cite{Hasenfratz:1989pk}
\be
 g_1 = 
 \frac{1}{(4\pi)^2} 
 \int_0^\infty \frac{{\rm d}\lambda}{\lambda^2} e^{-\lambda M_P^2}
 \sum_{n \in \mbox{\raisebox{-0.2ex}
 {$\scriptstyle\mathsf{Z\hspace*{-1.3mm}Z}^4$}}} 
 \Bigl(1 - \delta^{(4)}_{n,0} \Bigr)
 \exp \Bigl[
 -\frac{1}{4\lambda} \Bigl( 
 T^2 n_0^2 + L^2 \sum_{i = 1}^3
 n_i^2 \Bigr) 
 \Bigr]\; ,
\ee
and, in dimensional regularization \footnote{%
  The divergence of $G(M^2)$ for $d\approx 4$ cancels against
  those in the $\alpha_i$'s \cite{Gasser:1983yg}.},
\be
 G(M^2) = \int \frac{{\rm d}^d p}{(2\pi)^d} \frac{1}{p^2 + M^2}
 \; .  
\ee
In the quenched approximation (to the extent that it is well defined for this 
observable), these predictions are modified to be
\be
  M_P^V =  M_P 
 \biggl[
 1 + \frac{1}{2 F_P^2}
 \biggl(   
 \frac{\alpha}{2 N_c} + \frac{\alpha M_P^2 - m_0^2}{2 N_c}
 \frac{{\rm d}}{{\rm d} M_P^2} 
 \biggr) g_1 \,
 \biggr] 
 \;,\label{eq:quench1}
\ee
with the infinite-volume mass $M_P$ given by
\be
 M_P  =  M \biggl[
 1 + \frac{1}{2 F^2} \biggl( 
 \frac{\alpha}{2 N_c} G(M^2) + \frac{m_0^2 - \alpha M^2}{2 N_c} H(M^2)
 \biggr) - \frac{M^2}{2(4\pi F)^2}(\alpha_5 - 2 \alpha_8) \biggr] 
 \; . \label{eq:MPq}
\ee
In Eqs.~(\ref{eq:quench1}) and (\ref{eq:MPq}),
$\alpha/2 N_c$ and $m_0^2/2N_c$ are the parameters related
to the flavour singlet field (with the normalisation conventions 
of \cite{Hernandez:2002ds,Damgaard:2002qe}), and 
\be
 H(M^2) = \int \frac{{\rm d}^d p}{(2\pi)^d} \frac{1}{(p^2 + M^2)^2}
 \;.  
\ee

\end{document}